# Tuning Magnetic Properties Polycrystalline of PtCo Alloys Films with Pt

M. Erkovan[1] M Türksoy Öcal and O. Öztürk*[,2]

[1] *Sakarya University, Metallurgy and Material Science Engineering Dept., 54187, Serdivan, Sakarya-Turkey.*

[2] *Gebze Institute of Technology, Physics dept., 41400, Çayırova, Kocaeli-Turkey.*

**Abstract**

We experimentally investigated disordered $Pt_xCo_{1-x}$ (here x: 0.4, 0.5 and 0.6) alloy thin films magnetic properties which depended on Pt content. The magnetic properties of PtCo films were described with two effects, one of them is the hybridization between Co 3d and Pt 5d energy levels and it causes Pt magnetic polarization. The second one is the high spin orbit coupling constant of Pt which increases the ratio of magnetic orbital moment to spin moment. We investigated magnetic properties considering these effects by vibrating sample magnetometer (VSM) and ferromagnetic resonance (FMR) techniques.

**Introduction**

Controlling of magnetic properties of the films is the most important part in spintronic applications. This controlling is possible with the film structural form (magnetic multilayer, epitaxial film) and alloy form (include two or more elements). Alloy form in these structures is very critical especially for high density data storage applications. If you need keeping your data for a long time in a data storage media, you have to use large magnetocrystalline anisotropic magnetic materials [1-3]

Magnetocrystalline anisotropy constant value can tune with alloy form films. For example, the room temperature ferromagnetic materials Co, Fe and Ni have 410 $KJ \cdot m^{-3}$, 48 $KJ \cdot m^{-3}$ and 5 $KJ \cdot m^{-3}$ respectively [4]. As Co makes alloy with different elements such as Sm and Pt, its magnetocrystalline anisotropy value can be modified. The magnetocrystalline anisotropy value of $Co_5Sm$ is 17.2 $MJ \cdot m^{-3}$ [14, 5]. It is used nowadays in data storage media applications [5] and is forty thousand times larger than the pure cobalt's value. The other good example for these kinds of alloy forms is PtCo and it is promising for the next generation data storage media [6, 7] because of high magnetocrystalline anisotropy value (4,9 MJ·m$^{-3}$) of L1$_0$ phase Pt$_{0.5}$Co$_{0.5}$ form [6]. PtCo alloys have three well defined ordered phases: CoPt with the tetragonal L1$_0$ structure and both Pt$_3$Co and PtCo$_3$ with the cubic L1$_2$ phase [8]. There are a lot of studies on their ordered crystal structure [8-11] and magnetic properties but till now nobody has investigated their magnetic properties of polycrystalline forms in details. They can be prepared in disordered structure as a polycrystalline form because of their phase diagram. PtCo films also have good tolerance against to oxidation, corrosion and large Kerr rotations at short wave lengths [12-17]. Nowadays, their polycrystalline forms have been also attractive subject for a part of exchange bias film systems as a ferromagnetic layer [18-19]. Besides their magnetic usage area, their catalytic properties and their electrical properties are also active research subject [20-24]

The polycrystalline of PtCo alloys preparation steps and their magnetic properties are the subject of this study. Different chemical ratio Pt$_x$Co$_{1-x}$ (x=0.4, 0.5 and 0.6) alloy films were prepared in magnetron sputtering system from pure Pt and pure Co targets with co-sputtering technique. In literature, different preparation techniques have been used such as sputter deposition directly from PtCo alloy target [18] and Pt chips mounted in a Co target [25]. The chemical ratios of the PtCo films in this study were characterized by X ray photoelectron spectroscopy (XPS). Both vibrating sample magnetometer (VSM) and ferromagnetic resonance (FMR) techniques were used to investigate the magnetic properties of the films. This current study is also the first FMR investigation for polycrystalline PtCo alloy films. Curie temperature of the films also was specified with temperature dependent VSM measurements. Our findings are consistent with the previous studies of ordered PtCo alloy films.

**Sample Preparation and characterization**

The samples were fabricated and characterized in a cluster chamber combined with magnetron sputtering deposition chamber and analytical chamber. The films were grown onto naturally oxidized p-type Si (001) substrates by magnetron sputtering with base pressure <1x10$^{-8}$ mbar. All substrates were subjected to two cleaning process before they loaded in the vacuum chamber. The first one was done outside with ethanol and methanol bath. The second one was at UHV conditions they were annealed up to 600 °C to remove surface roughness and some dirt. The Pyrolytic Boron Nitride heater was used, at the sample holder, located under the substrates has the capability of annealing up to 1000 °C. The sample holder was cooled by chilled water to stabilize the sample temperature during annealing process. After annealing process, qualities of the substrates were checked with XPS.

Ar process gas (of 6N purity) was exposed to deposition chamber for magnetron sputtering process. The base pressure was increased from $1 \times 10^{-8}$ mbar to $1.2-1.4 \times 10^{-3}$ mbar level with argon exposing. PtCo alloy films were prepared with co-sputtering technique that both Co (3N5) and Pt (4N) were sputtered from elemental targets. The distance between the substrate and target can be varied between 10 mm and 100 mm but we used the largest distance for all growth process.

Quartz Crystal Monitoring (QCM) thickness monitor was used to observed deposition ratio in situ. It needs some parameters to measure the deposition ratio such as impedance value, the density of the deposition material and tooling factor. Especially the tooling factor changes from material to material, the most critical point in QCM is to measure the tooling factor correctly. Due to this reason, a QCM system needs some extra experimental results such as XPS, thin films XRD, etc. XPS is the one of the best techniques to calibrate the QCM parameter. The measurement results of QCM are calibrated by photoemission attenuations.

In order to determine the deposition ratios of the Pt and Co, the pure silver substrate was used to observe the attenuations of silver atoms photoemission as a function of additional material onto the substrate. For the thickness determination from XPS signal, the major photoemission peak of substrate and exposure material were used and their intensity ratios converted to the film thickness with electron mean free path value in the TPP (Tanuma- Powell-Penn) formula [26]. In this study, we used Ag 3d5/2 attenuation as a function of (Pt 4f7/2) and Co (Co 2p3/2) exposure. Co-sputtering deposition was used for PtCo alloys which were grown by repeated deposition sequences of sub-monolayers of Pt and Co. The number of sequences was changed according to the chemical ratio of the films. The shutters in front of each gun were opened during each sequence of depositions as the substrate on the holder was moved under certain gun of Co or Pt. The power applied for Pt target (DC gun,) and Co target (RF gun) were 2 watt and 25 watt, respectively. The deposition ratios with these powers were measured as 0.1 Å/sec for Pt and 0.3 Å/sec for Co. Based on deposition rates for every elemental deposition, the number of atoms of Pt and Co for each alloy type can be calculated per unit area per time. According to the elemental stoichiometry, the deposition periods of Pt and Co targets for each sequence are given in Table 1.

**Table 1.** Deposition periods for Pt and Co targets.

| Alloy | Pt deposition time (sec) | Co deposition time (sec) |
|---|---|---|
| $Pt_{0.4}Co_{0.6}$ | 4 | 4 |
| $Pt_{0.5}Co_{0.5}$ | 5 | 3 |
| $Pt_{0.6}Co_{0.4}$ | 6 | 2 |

During co-sputtering deposition, both Pt and Co targets were operated at the same time and the temperature of the substrate was 350°C. The chemical stoichiometry of the samples was characterized by using XPS in situ. The XPS spectra from alloy surfaces of $Pt_{0.4}Co_{0.6}$ sample were given in Figure-1.

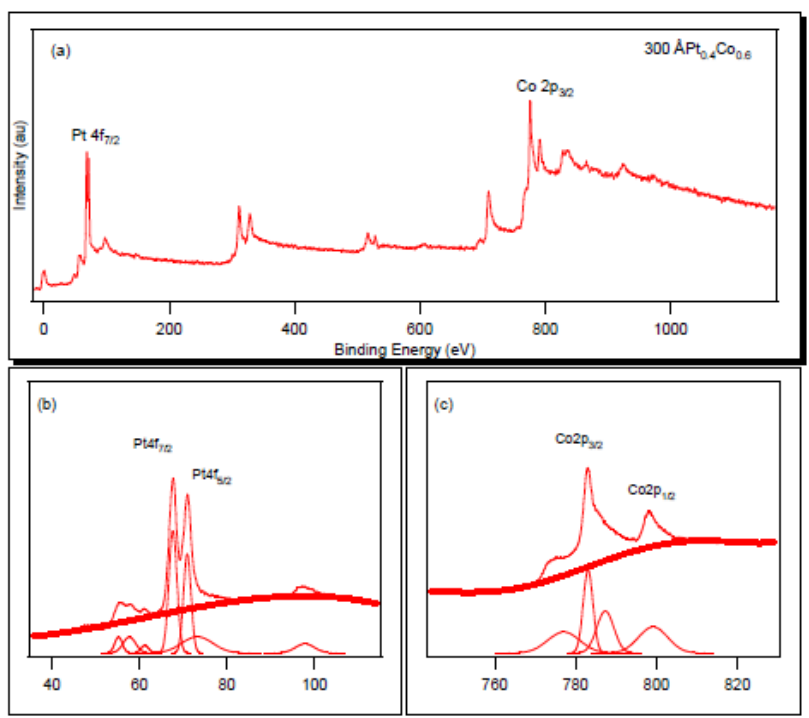

Figure-1: (a) XPS survey spectrum taken from $Pt_{0.4}Co_{0.6}$ (b) The high resolution spectra for the major peak of Pt and (c) The high resolution spectra for the major peak of Co.

In order to calculate Pt-Co ratios in the PtCo alloy, high resolution window XPS spectra should be taken from the major photoemission Pt 4f and Co 2p regions. Both Pt 4f and Co 2p peak integrated areas were calculated by CASA XPS 2.3.14 commercial software (SPEC GmbH). The Shirley background function was used for fit analyses of peaks. For the peak area calculation Voight function was used because it matches well with photoemission nature. After the peak areas calculation with this method, they were divided by atomic sensitivity factors which changes from element to element. The calculated Pt to Co ratios in PtCo alloy layers of the three samples are 0.4:0.6, 0.5:0.5 and 0.6:0.4.

**Results and Discussion**

Magnetic properties of the samples were investigated by VSM and FMR techniques. VSM Quantum Design PPMS 9T was used for the magnetization measurements at room temperature in plane geometry IPG; field parallel to the sample plane. The temperature dependence of magnetization measurements were performed from room temperature to 1200 K. The FMR measurements were carried out by using a Bruker EMX model X-band electron spin resonance spectrometer at microwave frequency of 9.5 GHz. The measurements were taken as a function of the angle of the external dc field with respect to the film normal at room temperature. Figure-2 shows the hysteresis loops of the samples. It is clear that in Figure-2 the saturation magnetization ($M_s$), the remanence field ($M_r$) and coercive field ($H_c$) values decrease with increasing Pt ratio in PtCo alloy films (Table-2). One can clearly see that as platinum content increases hysteresis loop gets narrower.

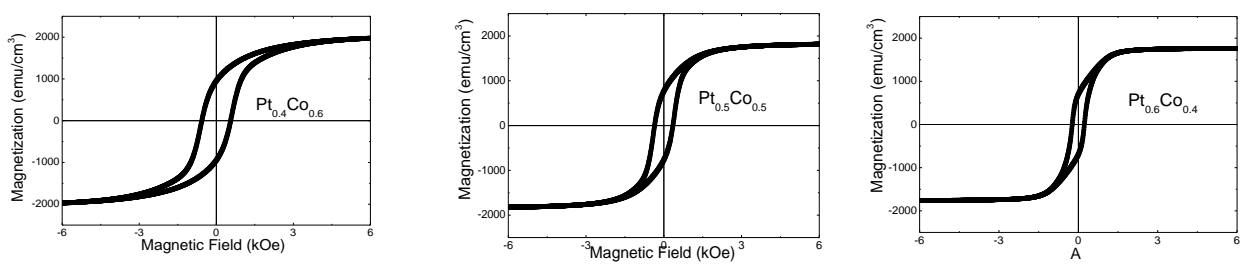

Figure-2: Hysteresis loops of samples were measured with VSM.

Table-2: Chemical ratio dependent of magnetic parameters.

| Alloy | $M_r$ (emu/cm$^3$) | $M_s$ (emu/cm$^3$) | $H_c$ (Oe) |
|---|---|---|---|
| $Pt_{0.4}Co_{0.6}$ | 943.621 | 2012.749 | 556.474 |
| $Pt_{0.5}Co_{0.5}$ | 758.944 | 1831.474 | 354.332 |
| $Pt_{0.6}Co_{0.4}$ | 694.023 | 1759.760 | 227.514 |

We noticed that change in the saturation magnetization, the remenance field and the coercive field values of samples didn't depend on decreasing Co ratio function linearly , It notes also that the samples magnetic behavior of the samples were shifted from hard to soft magnetic phases with increasing Pt contents in PtCo films.

Magnetic properties of the samples were also investigated by FMR technique. Figure-3(a) shows the schematic illustration of for relative orientation of the equilibrium magnetization vector M, the applied dc magnetic field vector H, and the experimental coordinate system. We observed two peaks from their FMR spectrum, Figure-3(b). One of them in low resonance field was called surface mode, the other one in high resonance field was bulk mode.

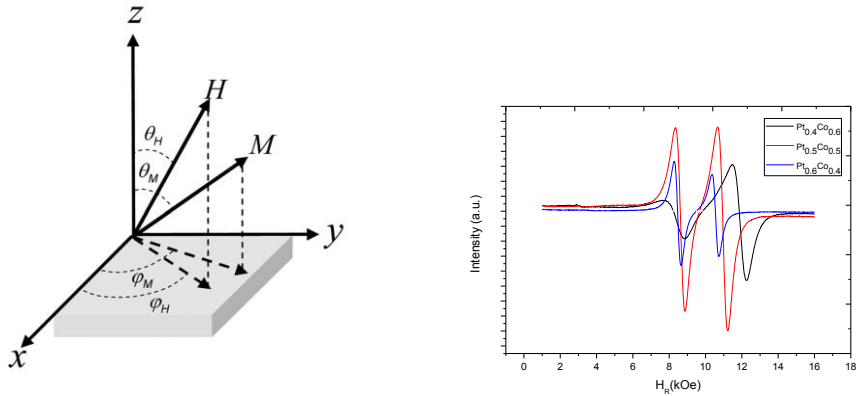

Figure-3: (a) Relative orientations of the external dc magnetic field and magnetization vectors with respect to the sample plane, (b) FMR spectra of $Pt_xCo_{1-x}$ alloy films for the external field perpendicular to the sample plane.

The angular variations of resonance fields for all samples are given in Figure-4. As one can see from the graph there is only one FMR mode until the magnetic field direction is nearly 10° away from film normal for all cases. When the field is applied very close to the film normal, additional FMR mode (so-called surface mode) appeared with lower resonance field values. The angular dependence of the curve of resonance field for the acoustic mode for both samples is almost the same for a broad range of angles. While the lowest resonance field values observed from $Pt_{0.6}Co_{0.4}$ sample, the largest resonance field values observed from $Pt_{0.4}Co_{0.6}$.

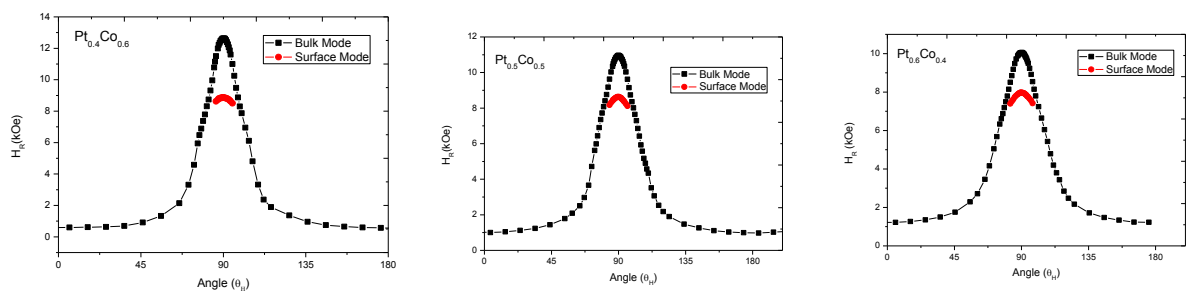

Figure- 4: Angle dependence of the resonance field values for all samples.

According to both VSM and FMR measurement results, we can explain the magnetic properties of the films with two possible mechanisms. The first one is the hybridization between Co 3d and Pt 5d orbital and it depends on the film composition. While the hybridization causes Pt to gain magnetic moment, spin magnetic moment of Co decreases with this effect. If the hybridization gets weak and weak between Co and Pt, the drastic decrease may be observed from Pt 5d ferromagnetic polarization while the magnetic moments of Co atoms enhance its anisotropic character [27]. This behavior was observed from our FMR results as Pt concentration increases, the Co magnetic anisotropy decreases. Similar hybridization effect observed from NiPt films [28-29] and magnetic moment for Ni in the NiPt alloy was measured with XMCD technique and the value was less than the magnetic moment of pure Ni's value. They also observed the spin polarization of Pt. [30]

The second one is the Pt's high spin orbit coupling constant. Pt has relatively large spin orbit coupling constant in 5d elements. If Pt and Co spin orbit coupling constants are compared, Pt has ten times larger value than Co [31]. Its effect is observed as Pt makes alloy with Co, Pt's high spin orbit coupling is transferred to the system. Therefore cobalt atoms are exposed to this large spin orbit coupling which is much weaker than exchange field. The large spin orbit coupling increases magnetic orbital momentum of cobalt in PtCo alloy. Gamberdalla et al [27] showed that Co gained huge magnetic orbital moment as Co atoms deposited on Pt (111). The ratio of magnetic orbital moment to spin moment is 0.099 [32] in pure Co on the other hand the ratio was determined as 0.16 [33] from disordered $Pt_{0.2}Co_{0.8}$. The value for disordered $Pt_{0.75}Co_{0.25}$ alloy was calculated as 0.21 [34]. Ordered $L1_0$ PtCo alloy films have 0.15 [35]. These results shows that disordered PtCo films may have higher ratio as Pt increases in PtCo alloy films. Co gains some magnetic orbital moment due to Pt's high spin orbit coupling value.

It is well known [36] that if the ratio of the orbital spin magnetic moment increases, the splitting factor g value reduces and also according to J. Pelzl et al., [37], there is a close relation between g value and Gilbert (G) damping parameter equation, *G* should be proportional to $(g - 2)^2$ [38]. In the light of this information, As Pt content increases, the ratio of the orbital to spin magnetic moment increases and it induces low g value and Gilbert damping parameter. If Gilbert damping parameter changes in any ferromagnetic system, this change can be observed from FMR spectra, the line width of peaks reduces, and the ferromagnetic material properties changes from hard or soft [36], and our VSM results showed similar behavior too. As Pt content increased in our samples the line width and the resonance field value reduced in FMR spectra.

If PtCo has ordered structure and it is in $L1_0$, perpendicular magnetic orientation is expected to be observed in these samples. We also checked it with FMR and VSM magnetometer; our samples do not have perpendicular magnetic orientation.

In this study we also measured the Curie temperature of the PtCo by temperature dependence VSM measurements. Temperature dependent magnetization measurement results are given in Table-3. Their Curie temperature reduces with increasing Pt ratio in PtCo alloys. According to Woodbridge [39] as the cobalt ratio increases in $Pt_xCo_{1-x}$ alloys the Curie temperature decreased with decreasing with Co ratio because of the diminishing of the interaction between neighboring spins. In another study, C Eyrich et al., [40] investigated change in their Curie temperature with both Density Functional Theory and X Ray Magnetic Circular Dichorism technique; their results are consistent with our results.

Table-3: The Curie temperature of Co and $Pt_xCo_{1-x}$ alloys.

| Sample | Curie Temp. (K) |
| --- | --- |
| Co | 1200 |
| $Pt_{0.4}Co_{0.6}$ | 970 |
| $Pt_{0.5}Co_{0.5}$ | 945 |
| $Pt_{0.6}Co_{0.4}$ | 915 |

**Summary**

The results of this study explained the role of Pt in different chemical ratio of PtCo alloy films. This role is related with Pt's high spin orbit coupling and hybridization with Co. In order to investigate these effect two different magnetic measurement techniques VSM and FMR were used. The FMR and VSM results showed that Pt's high spin orbit coupling and hybridization between Pt 5d and Co 3d affected the ratio of magnetic orbital moment to spin magnetic moment and this follows the changes in the magnetic properties for Co. Finally, Pt is non-magnetic element but because of hybridization it gains net magnetic moment in PtCo alloy films.

**Acknowledgements**

This work was supported by TUBİTAK (The Scientific and Technological Research Council of Turkey) through project numbers 106T576 and 212T217 and DPT project number 2003K120540.